\documentclass[conference]{IEEEtran}
\ifCLASSINFOpdf
  \usepackage[pdftex]{graphicx}
  \usepackage{listings}
\else
\fi
\usepackage{url}

\begin{document}
%
\title{Towards observability of scientific applications}



%
\author{\IEEEauthorblockN{Bartosz Balis\IEEEauthorrefmark{1}\IEEEauthorrefmark{2},
Konrad Czerepak\IEEEauthorrefmark{2},
Albert Kuzma\IEEEauthorrefmark{1}, 
Jan Meizner\IEEEauthorrefmark{1} and
Lukasz Wronski\IEEEauthorrefmark{1}\IEEEauthorrefmark{2}
}
\IEEEauthorblockA{\IEEEauthorrefmark{1}Sano Centre for Computational Medicine\\
Email: \{b.balis,k.czerepak\}@sanoscience.org}
\IEEEauthorblockA{\IEEEauthorrefmark{2}AGH Universtity of Krakow, Faculty of Computer Science, Poland\\
Email: balis@agh.edu.pl}
}


\maketitle

\begin{abstract}
As software systems increase in complexity, conventional monitoring methods struggle to provide a comprehensive overview or identify performance issues, often missing unexpected problems. Observability, however, offers a holistic approach, providing methods and tools that gather and analyze detailed telemetry data to uncover hidden issues. Originally developed for cloud-native systems, modern observability is less prevalent in scientific computing, particularly in HPC clusters, due to differences in application architecture, execution environments, and technology stacks. This paper proposes and evaluates an end-to-end observability solution tailored for scientific computing in HPC environments. We address several challenges, including collection of application-level metrics, instrumentation, context propagation, and tracing. We argue that typical dashboards with charts are not sufficient for advanced observability-driven analysis of scientific applications. Consequently, we propose a~different approach based on data analysis using DataFrames and a~Jupyter environment. The proposed solution is implemented and evaluated on two medical scientific pipelines running on an HPC cluster.
\end{abstract}


%
\IEEEpeerreviewmaketitle

\section{Introduction}
As software systems grow in size and complexity, conventional monitoring methods are becoming less effective in presenting a complete overview of the current state, or detecting and diagnosing performance issues in such systems. These traditional methods are limited to identifying anticipated problems, often overlooking unexpected issues or performance bottlenecks. In contrast, observability adopts a more holistic approach, using a suite of methodologies, strategies, and tools designed to gather and analyze detailed telemetry data. The primary goal of observability is to uncover unrecognized issues and investigate the "unknown unknowns" that plague the underlying computer systems and applications \cite{observability-definition}.

Modern observability emerged in the context of complex distributed systems, such as cloud-native applications built using microservice architectures \cite{kosinska2023towards}, or serverless systems \cite{cordingly2021enhancing}. Although scientific computing also relies on complex distributed systems, notably HPC clusters, the adoption of modern observability methods and tools in such systems has been slow for several reasons: 
\begin{itemize}
    \item Observability tools and libraries are typically designed for ``business'' applications which differ considerably from scientific applications in terms of architecture (services vs. jobs) and workload characterization (transactional vs. batch) 

    \item The execution environments also differ significantly. Business applications typically rely on cloud-native computing with Kubernetes as a workload manager. In such an environment, the application owner has a much stronger control over the observability stack than in HPC environments where batch job schedulers, such as SLURM, manage resource access.

    \item Consequently, technology stacks used in both systems are very different which strongly affects the applicability of observability tools and methods. Notably, scientific computing is not based on container orchestration, even though containerization is increasingly adopted in HPC systems \cite{cerin2020towards}. 
\end{itemize}

In this paper, we propose and evaluate an end-to-end solution to enhance observability within the realm of scientific computing. We identify the main challenges and propose an observability architecture for scientific applications in HPC environments that tackles them. We implement and evaluate the proposed architecture using a~medical scientific pipeline running on an HPC cluster.

The paper is organized as follows. Section \ref{sec:related} reviews related work. Section \ref{sec:problem} describes the research problem.  Section \ref{sec:solution} describes the proposed solution for observability of scientific applications. Section \ref{sec:exp} evaluates the solution. Finally, section \ref{sec:conc} concludes the paper.

\section{Related Work}

\label{sec:related}

In \cite{kunz2022hpc}, an architectural framework for job monitoring is proposed, leveraging the command from the SLURM workload manager to collect metrics, sending them to storage, identifying and tagging anomalies, and subsequently using platforms such as Grafana for alert generation and data visualization.

To extract SLURM metrics, the authors devised a bash script that fetches metric data via the \emph{sacct} command, allowing real-time extraction of job execution information. Prometheus, a time series database, was employed for the storage of this monitoring data.

For data visualization, the authors used Grafana, which facilitated a concise overview of ongoing jobs and associated anomalies. Utilizing Grafana's table visualization functionality, the authors were able to create custom tables with the flexibility to adjust column sizes, alignments, and column order, enabling a comprehensive and clear representation of job execution.

However, this work primarily focuses on metrics, not considering logs and traces from running jobs. Moreover, our approach based on data analysis tools offers much greater flexibility in terms of analysis and visualization capabilities. 
 
In \cite{chan2019resource}, an architecture to monitor a high-performance computing cluster utilizing freely accessible open-source tools such as InfluxDB, PostgreSQL, and Grafana is proposed. This configuration establishes a framework for data visualization:

\begin{itemize}
    \item Telegraf -- Deployed across all nodes within the cluster, this server agent is responsible for collecting, processing, and forwarding metrics. The 'exec' plugin incorporated within Telegraf runs a Bash script at preset intervals to accumulate SLURM job data, which is subsequently stored in InfluxDB.
    \item PostgreSQL -- This database was employed to support complex queries to extract the gathered data.
    \item Grafana -- This tool was employed for analyzing and gaining insights from the collected data.
\end{itemize}

The disadvantage of the solution is the requirement to run a Telegraf daemon on the nodes within the cluster. In contrast, our solution is based solely on the permissions granted to the users.

In \cite{guilbault2023self}, the authors emphasize the importance of job-level metrics. They utilize Prometheus and develop several exporters for SLURM-based HPC systems to gather these metrics. For visualization, they created a custom portal displaying various charts. However, this approach has several limitations. In our experience, Prometheus is well-suited for node-level monitoring but not as effective for job monitoring due to its strict pull model of data collection. This model complicates the collection of telemetry data generated by instrumentation code within application processes. Additionally, developing a custom portal offers less flexibility compared to our approach, which leverages data analysis in Jupyter.  

\section{Problem statement}
\label{sec:problem}
\begin{figure}[!htb]
    \includegraphics[width=0.4\textwidth]{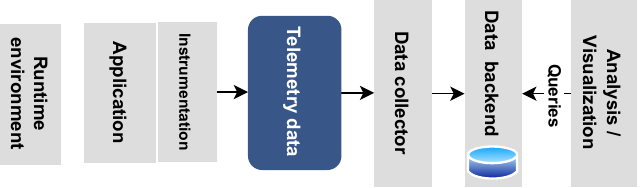}
    \centering
    \caption{General observability architecture}
    \label{fig:genarch}
\end{figure}

Large scale and complexity of modern scientific applications makes their observability an important concern. Fig. \ref{fig:genarch} presents a general architecture of a~modern software observability ecosystem. The \textit{Application} together with its (distributed) \textit{Runtime environment} is the system under observation. \textit{Telemetry data} -- metrics, traces, and logs -- are generated by (a) monitoring agents deployed in the runtine environment, and (b) instrumentation: additional code injected into the application.  Telemetry data are collected by a~\textit{Data collector} which delivers them to downstream components.  OpenTelemetry~\cite{opentelemetry-site} divides the data collector layer into Receivers (consuming data), Processors (transforming data), and Exporters (exporting data to backends). \textit{Data backend} is responsible for the long-term storage and provision of telemetry data. Various \textit{Analysis and visualization tools} query the data backend to perform analysis.

\begin{figure}[!htb]
    \includegraphics[width=0.4\textwidth]{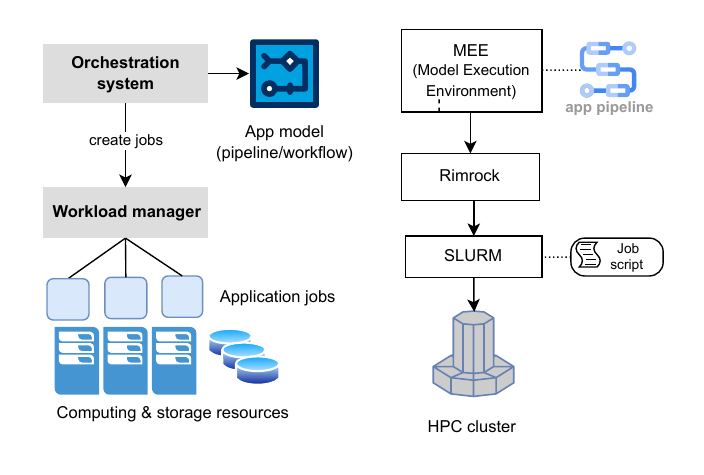}
    \centering
    \caption{Scientific application ecosystem: (left) abstract architecture; (right) specific implementation utilized in this study.}
    \label{fig:sciapparch}
\end{figure}

In this paper, we consider a general model of a~scientific application, shown in Fig. \ref{fig:sciapparch} (left), that consists of \textit{pipeline/workflow of jobs}. Such applications require an \textit{orchestration engine} which executes the pipeline by submitting jobs to a~\textit{workload manager} responsible for allocating access to resources of an underlying distributed computing infrastructure. While this general description fits such diverse systems as HPC clusters, cloud-based Kubernetes clusters, or serverless computing services, here we focus on the HPC systems, as they pose unique challenges for observability. The specific implementation of the architecture used in the experimental part of this study is depicted in Fig. \ref{fig:sciapparch} (right).   

The Model Execution Environment (MEE) \cite{meee-eur-valve} is a tool for executing scientific pipelines on high performance computing (HPC) systems. In this case, it is integrated with the Ares supercomputer hosted by the Cyfronet Computing Centre in Krakow. Rimrock is a~service that exposes SLURM job submission functionality as a~REST service. SLURM is a well-known cluster manager and job scheduling system used in HPC clusters.

The overall research goal can be broken down into the following detailed problems:

\begin{itemize}

    \item \textbf{Collection of application-level metrics}. Although system-level metrics, such as CPU and memory usage on cluster nodes, are usually available in HPC systems, application-level metrics give much more insight into the application behaviour. These include, for example, CPU and memory usage caused by the individual application jobs, their IO activity, network usage, etc. However, currently there is no standardized solution to collect such data in HPC systems.
    
    \item \textbf{Instrumentation} of distributed scientific application components to collect tracing data.

    \item \textbf{Context propagation} and correlation of telemetry data: developing robust methods for the effective transfer of telemetry data context, and correlation of telemetry data originating from distributed components.

    \item \textbf{Provision of monitoring agents} on compute nodes: this implies the deployment of monitoring agents on each compute node, allowing the collection of application-level telemetry data and its online transfer to the data collector. 



    \item \textbf{Analysis of telemetry data}: developing effective methods aiding in interpretability and accessibility of the data.
    \end{itemize}

\section{Solution for scientific applications observability}
\label{sec:solution}

\subsection{Proposed architecture}
\begin{figure}[!htb]
    \includegraphics[width=0.4\textwidth]{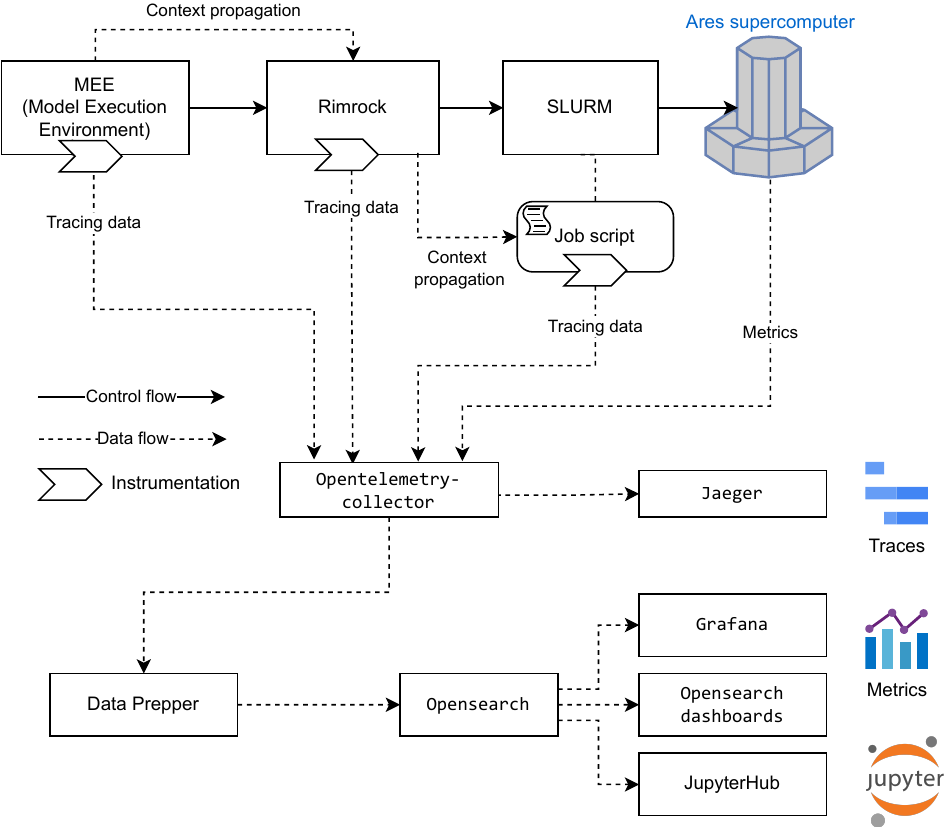}
    \centering
    \caption{Proposed observability framework architecture}
    \label{fig:arch}
\end{figure}

The main goal of the proposed architecture is to enable observability in the specific environment of HPC clusters. 
Fig. \ref{fig:arch} shows the implementation of the architecture for the execution environment based on MEE, Rimrock and SLURM (see section \ref{sec:problem}).

The architecture is based on open standards and open source software: OpenTelemetry, OpenSearch, Grafana, and JupyterHub. OpenTelemetry is a framework designed for collecting telemetry data (metrics, traces, and logs) from distributed systems \cite{opentelemetry-site}. OpenTelemetry delivers important mechanisms and associated software for, among others, manual and automatic instrumentation, context propagation, and a vendor-neutral Opentelemetry collector for data distribution.

The OpenTelemetry collector is a~vital part of the introduced architecture as it is responsible for receiving all telemetry data collected by the system and exporting it to downstream recipients, possibly also filtering, batching and processing it \cite{opentelemetry-collector-architecture}. Data prepper is an auxilliary component acting as a~bridge between the OpenTelemetry collector and OpenSearch. OpenSearch is a~distributed search engine, also providing database-like functions, that has been used for various data-intensive tasks, including analysis of logs and monitoring data \cite{papadopoulos2024architecting}. 

The implementation of the observability stack that allows its deployment using Docker compose is available at {\footnotesize \url{https://github.com/SanoScience/observability}}.

\subsection{Tracing of scientific applications}

Traces are important in an observability architecture, as they are responsible for gathering information about data flow in distributed systems. Trace analysis can provide information about the time spans in which a specific job was executed and its execution context.

To collect trace data, we have introduced instrumentation to three components: MEE, Rimrock, and, importantly, the Slurm job scripts. A~new trace (with a~unique \textit{traceID}) is created in MEE when a~new pipeline is started. The context data for this trace (e.g. the \textit{traceID}) is then propagated to Rimrock, and finally to the job script. This allows for the correlation of the telemetry data generated in the distributed components. 

To instrument MEE and Rimrock, we were partially able to utilize the auto-instrumentation features of OpenTelemetry. Instrumenting the jobs themselves was less trivial. Direct SLURM instrumentation would work best, but it would require changing the source code and deploying a custom SLURM build. Consequently, we have decided to instrument the SLURM job scripts which are essentially Linux shell scripts. To this end, we have prepared shell functions that utilize the OpenTelemetry tool, otel-cli,  which enables manual generation of traces using a~command line interface. 


\begin{figure}[!htb]
    \includegraphics[width=0.5\textwidth]{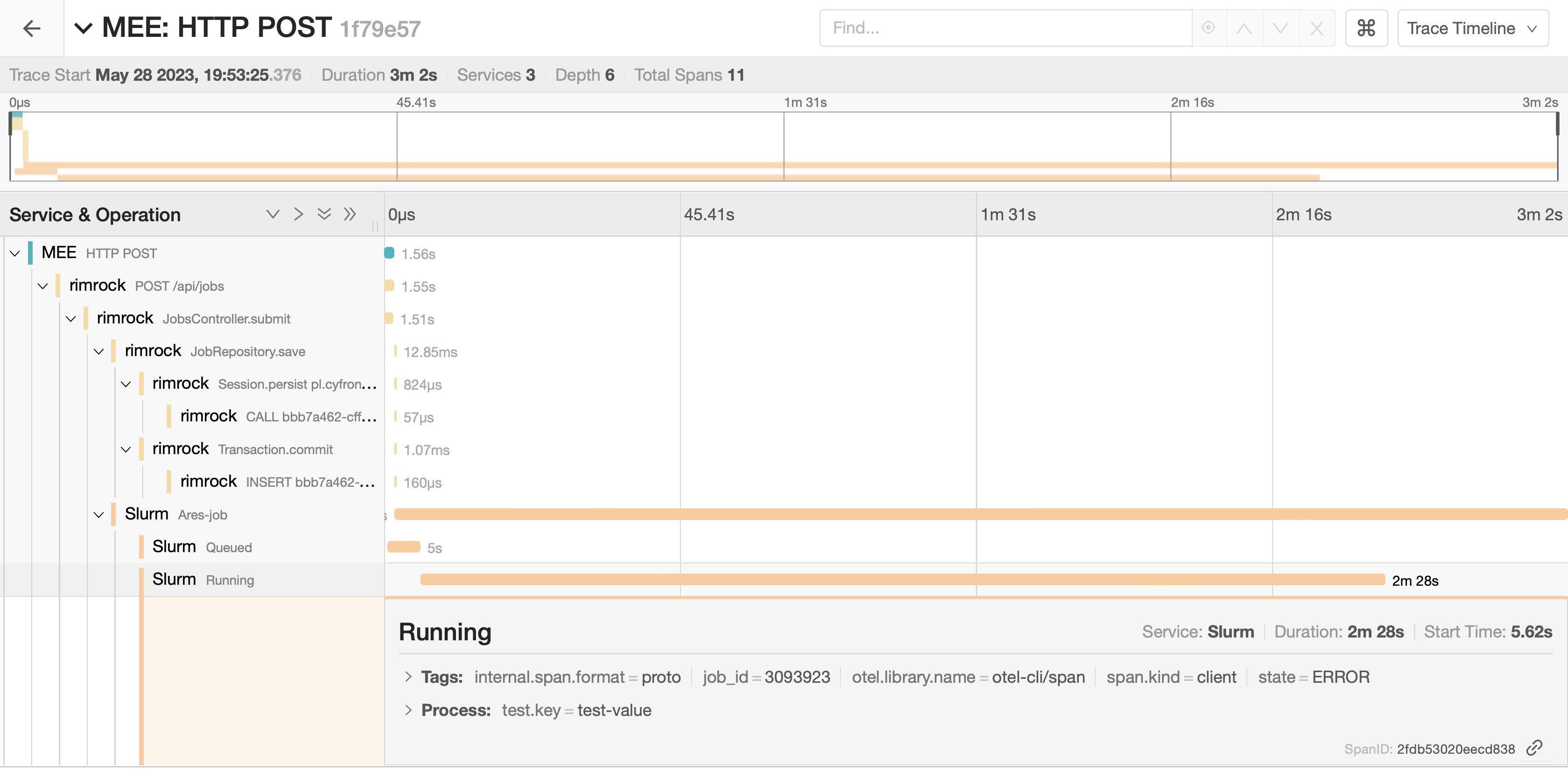}
    \centering
    \caption{Trace visualization in Jaeger.}
    \label{fig:trace_vis}
\end{figure}

Traces can be visualized in using a~number of tools, including Jaeger, an open source distributed tracing platform. An example trace visualization is shown in Fig. \ref{fig:trace_vis}. 

\subsection{Collecting application-level metrics}

The key data collected and stored by the observability architecture are metrics. Similarly to traces, we provide dedicated shell functions for collecting metrics that have to be injected into the script running the SLURM jobs. The commands download and run a~Python script in the background which sends metric data to the Opentelemetry collector periodically.

To collect metrics from a HPC machine, the \textit{cgroups} Linux kernel feature was used. Control Groups (cgroups) manage groups of processes running in the Linux kernel. It allows monitoring resource utilization of processes running in the same group. The key feature of cgroups in the case of monitoring jobs is the organization of processes into a hierarchical structure. Data about a~cgroup in which the monitored job is running provide the information needed for the observability architecture to extract metric data.

Control Groups are strictly integrated with Slurm, which makes data on memory and CPU usage accessible in the directory with path:
\textbf{/sys/fs/\-cgroup/\-memory/\-slurm/uid\_\{\}/job\_\{\}/}. 
The collected metrics include data on the job's memory usage, CPU usage, and the number of open and active files.

The metrics data is stored in the OpenSearch database. An auxiliary component provided by OpenSearch, Data Prepper, receives data from the OpenTelemetry collector, filters it, and passes accumulated metric data to Opensearch. The OpenSearch database exposes a~RESTful API to query data. This API is used both by the Data Prepper to send the metric data, and by the visualization/analysis tools to retrieve it.

\subsection{Custom telemetry metadata}
Rich telemetry metadata is crucial, as it allows one to put the telemetry data in appropriate context. Some metadata, e.g. trace and span identifiers, is passed automatically as context information. However, for the purpose of data analysis, it is desirable
to have the possibility to define custom, domain-specific metadata attributes. To this end, we propose a~mechanism of custom tags that can be passed in the job scripts when the monitoring is initialized. For example, one can use the following command: 





\begin{verbatim}
run_monitoring "--case-number $CASE_NUMBER 
 --pipeline-identifier $PIPELINE_IDENTIFIER 
 --pipeline-name $PIPELINE_NAME 
 --step-name $STEP_NAME"
\end{verbatim}

Here, \textit{case number}, \textit{pipeline identifier}, \textit{pipeline name}, and \textit{step name}, are all custom attributes that will be passed to the observability database. Consequently, when analyzing the data, the user will be able to filter or group by these metadata attributes. The values for the attributes are passed as environment variables which is a~portable and versatile mechanism.

\subsection{Visualization}

For metric visualization, we can use open source solutions, such as the Grafana or Opensearch dashboards. Such tools provide the user with an easily accessible set of dashboards with predefined charts. One advantage of dashboards is that they can be used in a~semi-online fashion, allowing to monitor jobs while they are running. 

\begin{figure}[!htb]
    \includegraphics[width=0.5\textwidth]{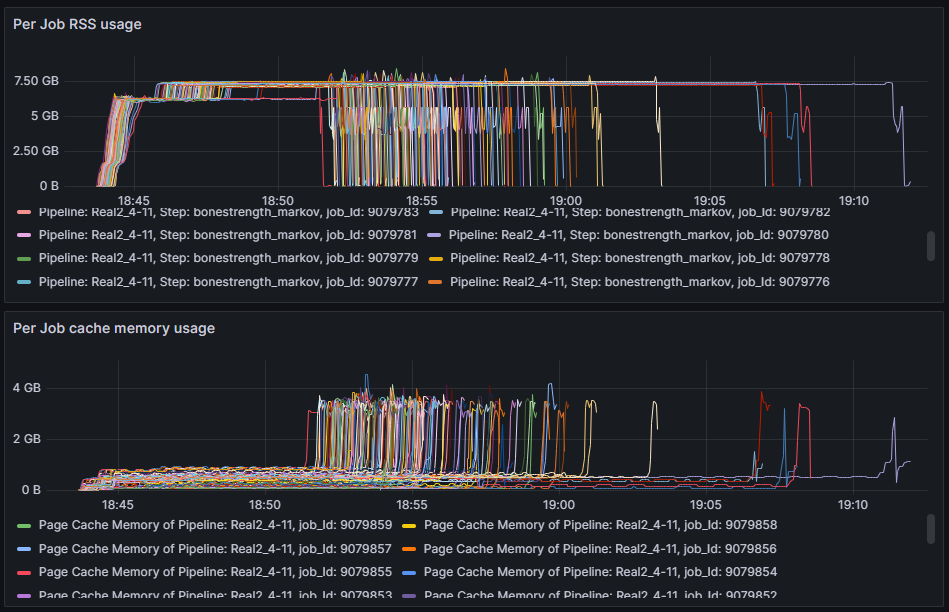}
    \centering
    \caption{Metric visualization in Grafana.}
    \label{fig:metric_vis}
\end{figure}

In Fig. \ref{fig:metric_vis}, an example Grafana dashboard is shown. All dashboards are interactive and can be modified by setting predefined filters, or selecting a~specific time period. However, it is not easy to add entirely new charts to existing dashboards. 

\subsection{Data analysis in Jupyter}

Although existing tools used in observability for visualizing metrics or traces, such as Grafana, OpenSearch dashboards, or Jaeger, are useful, they are not sufficiently flexible and customizable. For the purpose of performance analysis of scientific applications, more advanced data analysis capabilities are needed. 

Consequently, we have created an API to read telemetry data as Pandas DataFrames into a~Jupyter environment. We have also provided additional functions for creating typical visualizations, such as histograms. This approach is the cornerstone of our solution and is described in action in the next section.  

\section{Evaluation}
\label{sec:exp}

The solution is evaluated using two medical scientific applications: BoneStrength and AngioSupport.

The BoneStrength model from the \textit{In Silico World} project\footnote{In Silico World homepage: \url{https://insilico.world}}  follows the Digital Twin approach to represent an in-silico model of a hip bone \cite{kasztelnik2023digital}. The model utilizes the Finite Element Method (FEM)-based simulations backed by the ANSYS Mechanical suite to simulate bone load and its fracture during specific patient fall conditions. 
    
The monitored experiment constitutes part of the BoneStrength in-silico trial, which is a practical use case to estimate the risk of bone fracture within a set of patients (called the \textit{cohort}) in the form of a campaign simulation. A~single campaign simulates thousands of generated falls which are randomly assigned to the cohort's patients along a certain observation time. Processing a~single fall case (supported with Intel-MPI) requires 4 CPUs, at least 8GB of memory, and up to 5GB of disk space for the ANSYS software working directory. 

The AnigoSupport model predicts pressure and flow waves throughout the coronary arteries. It enables prediction of so-called Flow Fractional Reserve (FFR)~\cite{angio2019} after the surgical intervention. Traditionally, the FFR may be measured only with complex invasive procedure, yet it is crucial to assess the risk of Coronary Artery Disease (CAD) and influence the choice of treatment, such as a Percutaneous Coronary Intervention (PCI) or a Coronary Artery Bypass Graft (CABG), if needed. As this disease is claiming a~great number of lives worldwide, improvement of the diagnosis and treatment process is crucial from both ethical and economical point of view. Obviously, the ability to estimate the post-treatment FFR with a computational model processing the original images from the patient is greatly increasing the confidence of the decision on which course of treatment would be preferred and if it should  be carried out at all. While there are significant benefits of using the simulation-based solution it also is highly challenging due to its computational complexity. Existing High Performance Computing (HPC) systems made this type of solution possible, however neither the amount of HPC resources nor the time in which result is needed are infinite, hence performance monitoring and optimization are desirable. 

The AngioSupport model was used for running several simulations for one patient. In further charts metric data are shown for a~single patient workflow consisting of 5000 simulations divided between 300 workers. Workers are independent SLURM jobs that run a~python script which sequentially executes an assigned set of simulations in a~loop. In other words, each SLURM job executes many application-level `tasks' (simulations). To analyze performance at a~finer granularity (tasks), our observability solution leverages tracing and application instrumentation. We generate a new trace ID for each pipeline, and create spans for each SLURM job and its respective tasks. Additionally, a unique identifier is assigned to each task to facilitate detailed data analysis. 

Both described models are well-suited for evaluating the solution. On one hand, each MPI-based simulation demands significant computational resources; on the other, the large cohort size necessitates running multiple computations in parallel as SLURM array jobs. Additionally, the Model Execution Environment, featuring the monitoring platform discussed in this paper, is ideally suited for various models, enabling the execution of multi-step pipelines for multiple patients.

During the experiment, we successfully tracked the progress of array tasks in real time, allowing us to detect anomalies and identify simulation cases that significantly deviated from others. Additionally, we collected all necessary job metrics for a thorough analysis of resource consumption.

The metrics collected from the simulation runs were further analyzed using a specialized Jupyter notebook. To facilitate custom chart creation and data filtration, the script `read\_data.py` was integrated into the JupyterHub environment. This script enables the direct download of data from the OpenSearch service into a Pandas DataFrame. The provided starting notebook demonstrates how to use this script and includes some basic chart creation functions.



\begin{figure}[!htb]
    \includegraphics[width=0.5\textwidth]{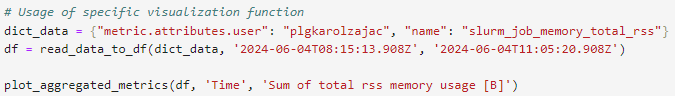}
    \centering
    \caption{Example code generating aggregation chart}
    \label{fig:code_example}
\end{figure}

To vizualize certain metrics, the user has to download metric data from the OpenSearch via shared script and use a prepared chart creating function, or create its own using the Pandas API and generally available Python libraries, such as matplotlib or seaborn. An example code for performing such a task is shown in Fig. \ref{fig:code_example}. By setting variables start\_time and end\_time the user defines a time period to download metric data. An important variable is the dict\_data, which enables filtering all metric data in the chosen period of time.

\begin{figure}[!htb]
    \includegraphics[width=0.5\textwidth]{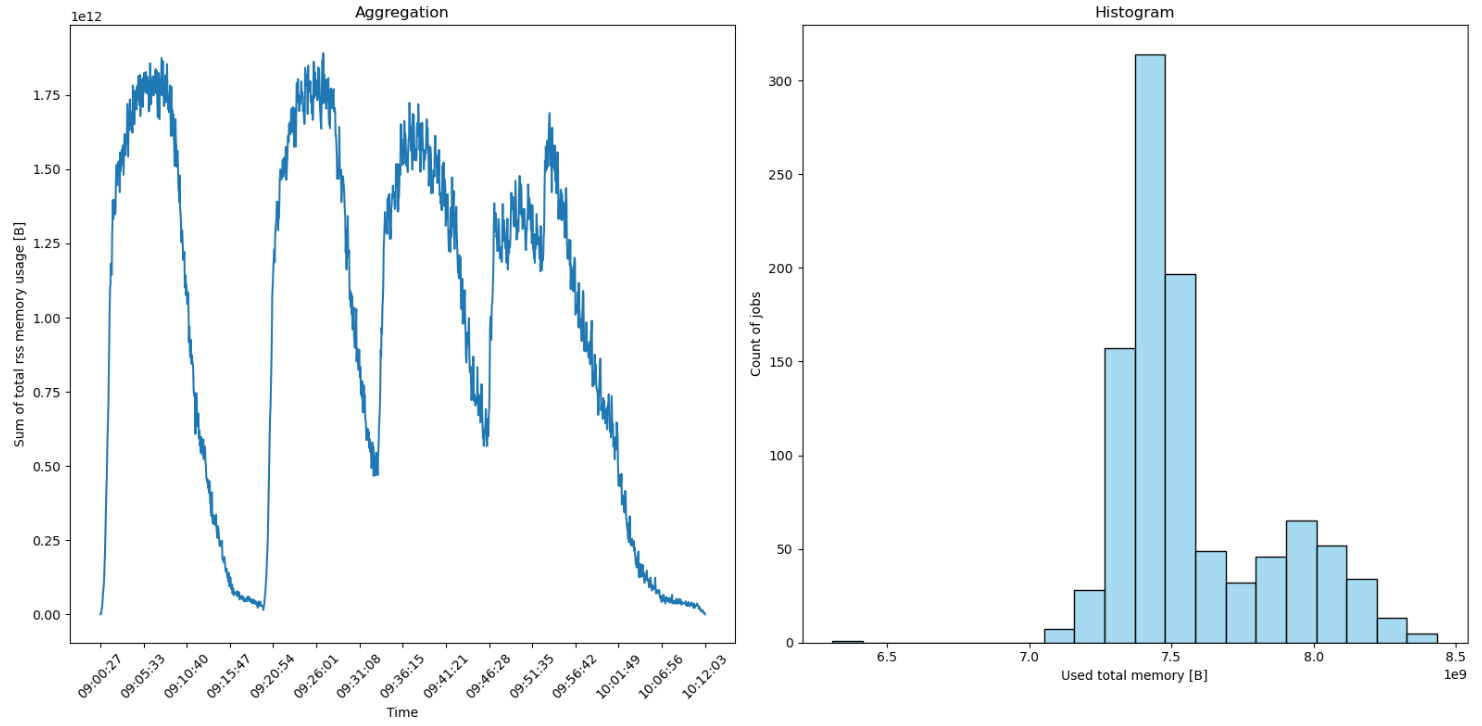}
    \centering
    \caption{Memory consumption: total usage during execution (left) and distribution of maximum memory usage by jobs (right).}
    \label{fig:RSS}
\end{figure}

The results for the most important metrics are presented as plots. They include total residental memory (shown in Fig.~\ref{fig:RSS}) which is a critical resource, as its exhaustion would crash the computations. 

 The distribution in Fig.~\ref{fig:RSS} reveals that the maximum memory consumption by jobs is quite uniform, varying between 7.1 and 8.4 GB. This enables us to select appropriate job size to utilize the HPC nodes to the fullest potential without risking overallocation that would result in abnormal termination of computations. 
 
 On the other hand, the total memory consumption during execution is quite variable. However keeping appropriately balanced set of jobs we can balance the whole computation to acquire best performance an utilization of the cluster.

\begin{figure}[!htb]
    \includegraphics[width=0.5\textwidth]{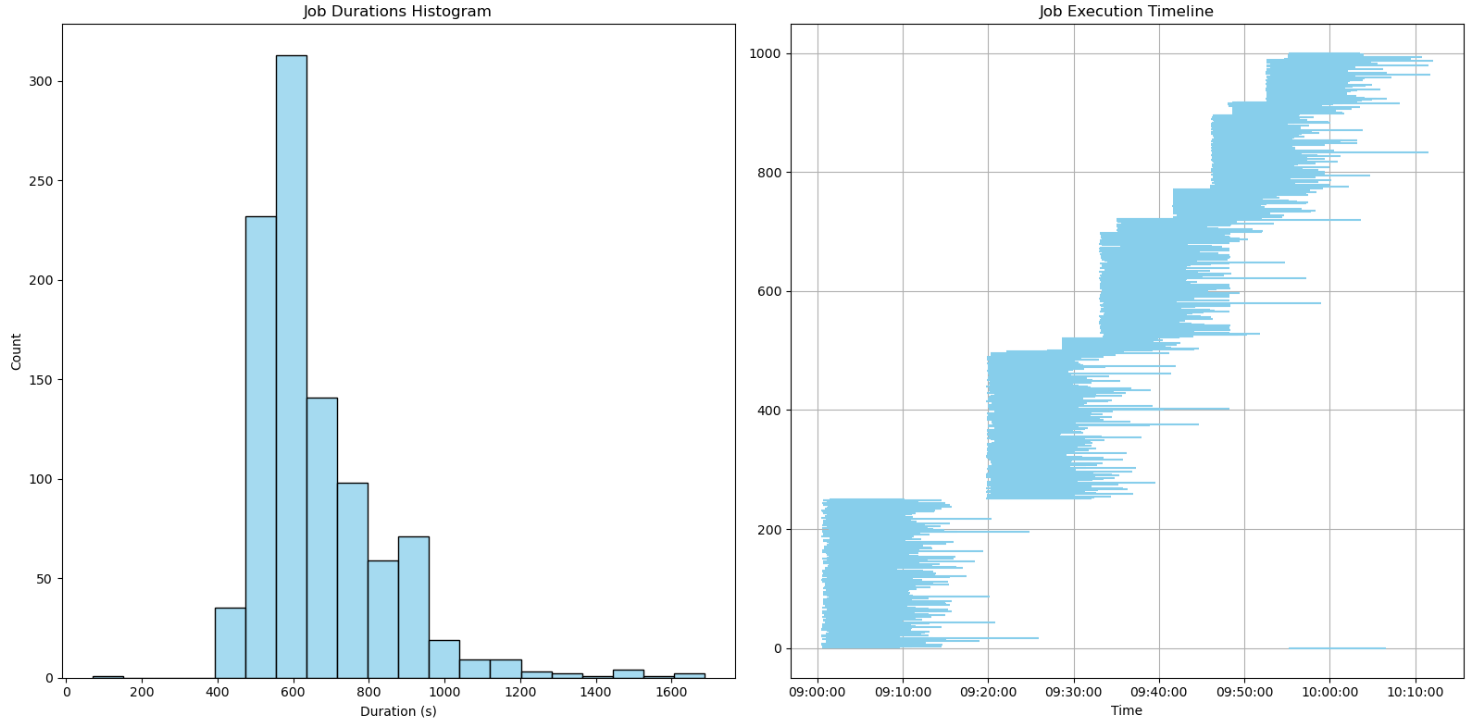}
    \centering
    \caption{Duration of jobs: distribution (left) and overall timeline (right). }
    \label{fig:timeline}
\end{figure}

\begin{figure}[!htb]
    \includegraphics[width=0.5\textwidth]{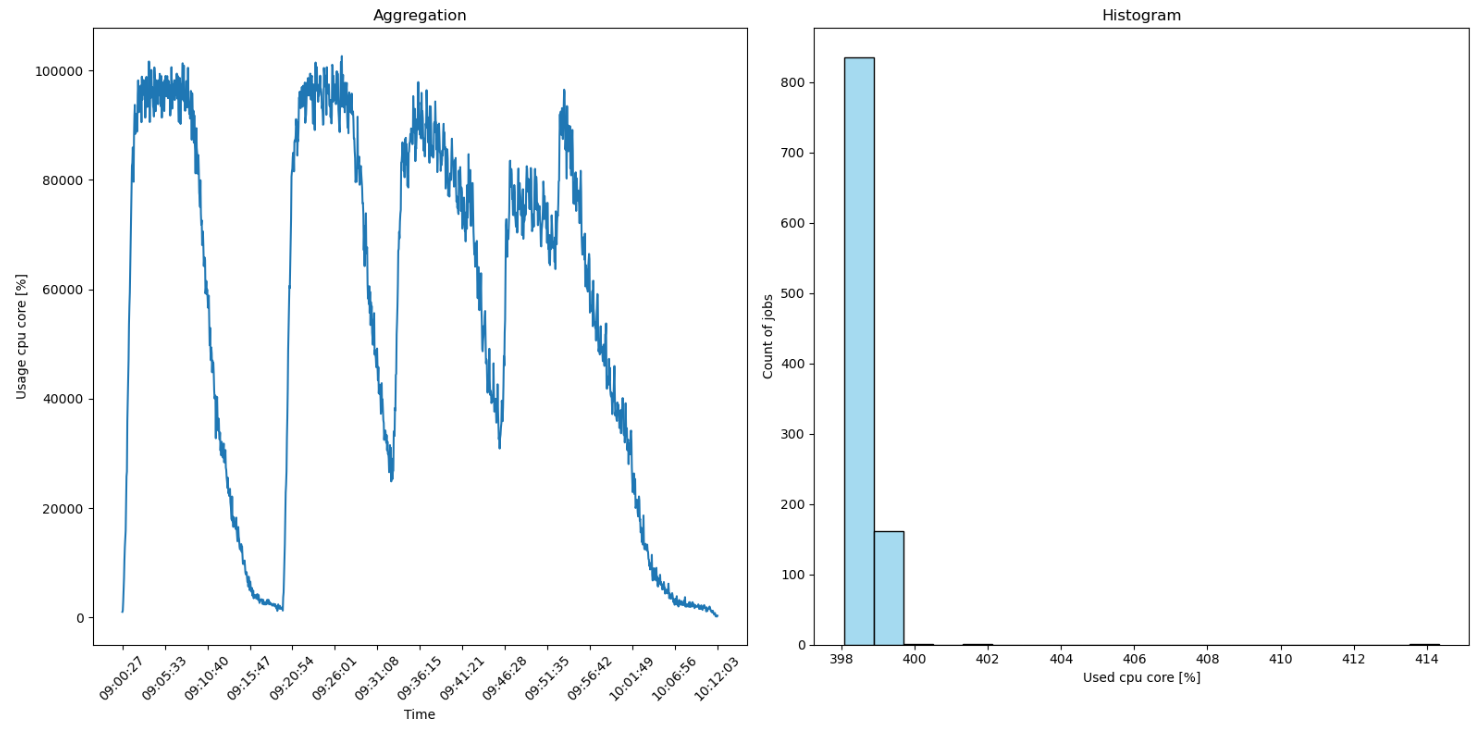}
    \centering
    \caption{CPU usage: total usage during execution (left) and distribution of maximum usage by jobs (right). All jobs are allocated 4 cpu cores and clearly they achieve this utilization. The overall usage is quite variable with large periodic peaks.}
    \label{fig:CPU}
\end{figure}

 Also the total CPU usage is shown in Fig.~\ref{fig:CPU} as its optimization is critical to enable running sufficient number of simulations in reasonable time. The distribution reveals a~``long tail'' of jobs with execution times much longer than most. The timeline confirms that around 250 jobs run in parallel which is the limit of available ANSYS licenses. While we do not control the policies of the HPC center such as the amount of provide licenses detection of such condition may allow us to plan for best utilization of the available nodes e.g. by co-scheduling different types of jobs that are free from such limitations such as pre- or post- processing jobs.

 Finally, metrics were also used to create the histogram depicting the relation between the number of jobs and their CPU usage, as shown in Fig.~\ref{fig:CPU}. It enables us to analyse the demand for computational power not just for a single patient but the whole cohort. By analyzing the chart, we observe that most jobs reach the maximum resource allocation of 4 cores, which is restricted by SLURM using the cgroups mechanism. This indicates that a higher allocation might potentially improve efficiency.

 The job duration timeline (shown in Fig.~\ref{fig:timeline}) highlights that the primary constraint on the experiment's elapsed time was the number of available ANSYS software licences (250) which is also the maximum number of jobs running in parallel. As some jobs fluctuated around the average execution time, jobs allowed new samples to move from the queue into the running state. 

 \begin{figure}[!htb]
    \includegraphics[width=0.5\textwidth]{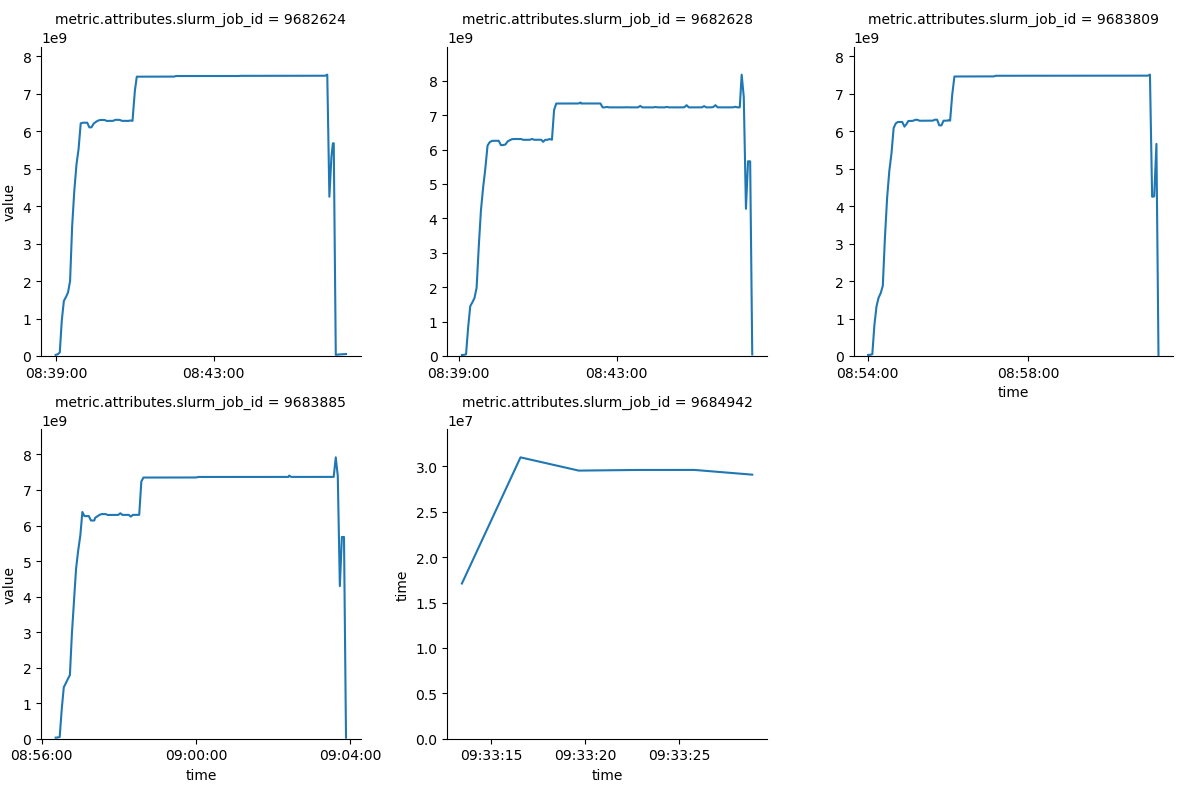}
    \centering
    \caption{ Grid plot for outlier simulation analisis. Charts show usage of RSS memory for five shortest simulations in whole set (timewise).}
    \label{fig:grid}
\end{figure}
 
 During simulation, some significant outliers occurred, indicating early errors in sample processing. Monitoring can help detect them, thanks to tracing data and find specific model parameters causing the issues within the huge parameter space. The prepared environment in JupyterHub empowers the user with filtering and visualizing tools for metric data. The functions prepared to generate visualizations per SLURM job enables analysis of shortest BoneStrength simulations (Fig.~\ref{fig:grid}). Receiving such detailed data on the simulation performance allowed for the detection of single slurm job which lasted only a couple of seconds. By investigating logs for the simulation, errors in the provided input file were detected.

 It is noteworthy that the ANSYS software or the model itself provides its own logs and metrics, e.g. from the MPI processes. Future work includes integrating monitoring data produced by the application, in addition to the collected SLURM job data, to enhance the workflow control and observability.

The second use case tested is the AngioSupport. It is also backed by the numerical simulation however as its nature differs it seemed reasonable to see what bearing it has on the performance of the solution. 

\begin{figure}[!htb]
    \includegraphics[width=0.5\textwidth]{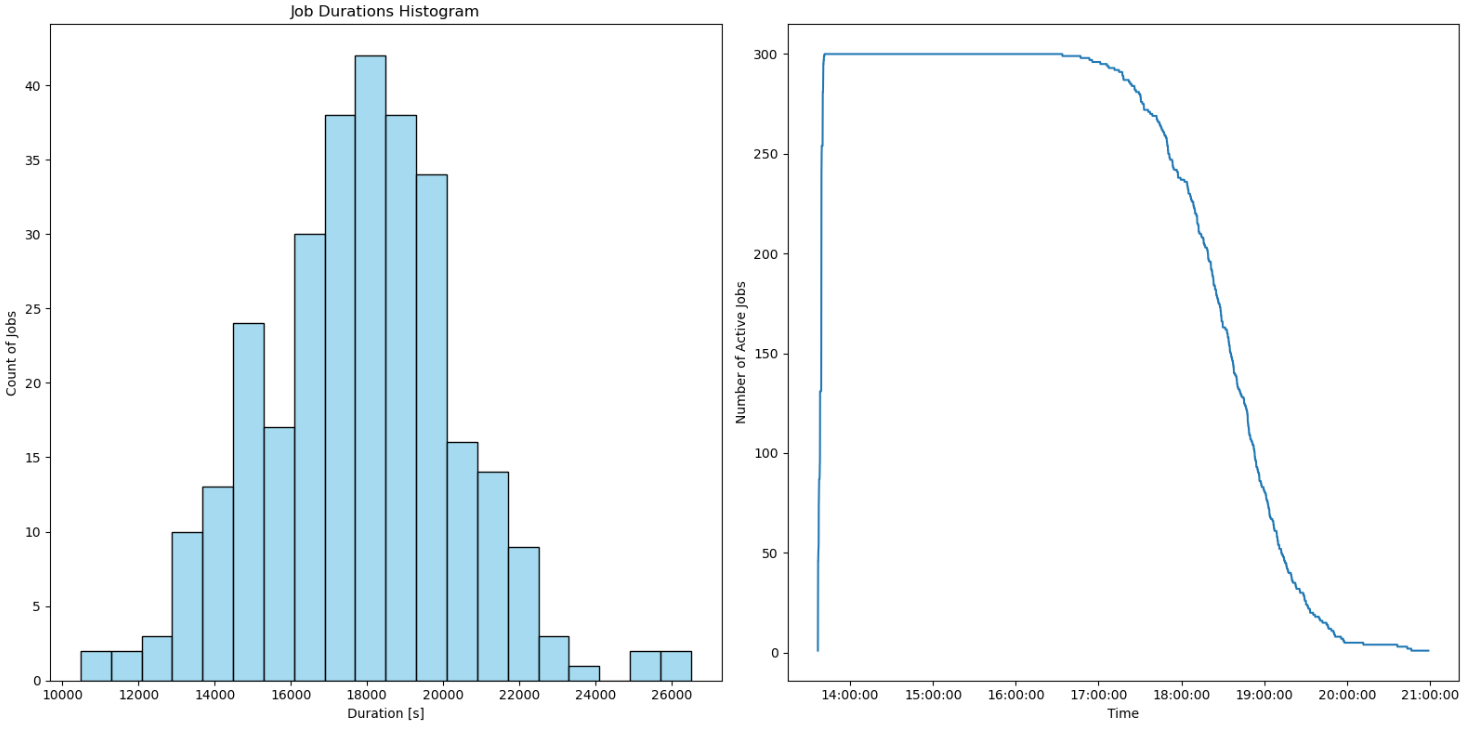}
    \centering
    \caption{ AngioSupport jobs length distribution (left) and timeline showing number of active jobs per time (right) }

    \label{fig:job_len}
\end{figure}

The first measurement in Fig.~\ref{fig:job_len} depicts jobs length distribution (left) and timeline showing number of active jobs per time (right). 

Both values are highly important to plan the most efficient way to plan the campaign run. As we can see most jobs take between 16 000 and 20 000 seconds – we can also organize shorter runs to fill spaces left by the longer ones.

Like in case of BoneStrength the total residential memory (RSS) was measured. This value is important to plan the computation in a way that would not cause it to crash.

\begin{figure}[!htb]
    \includegraphics[width=0.5\textwidth]{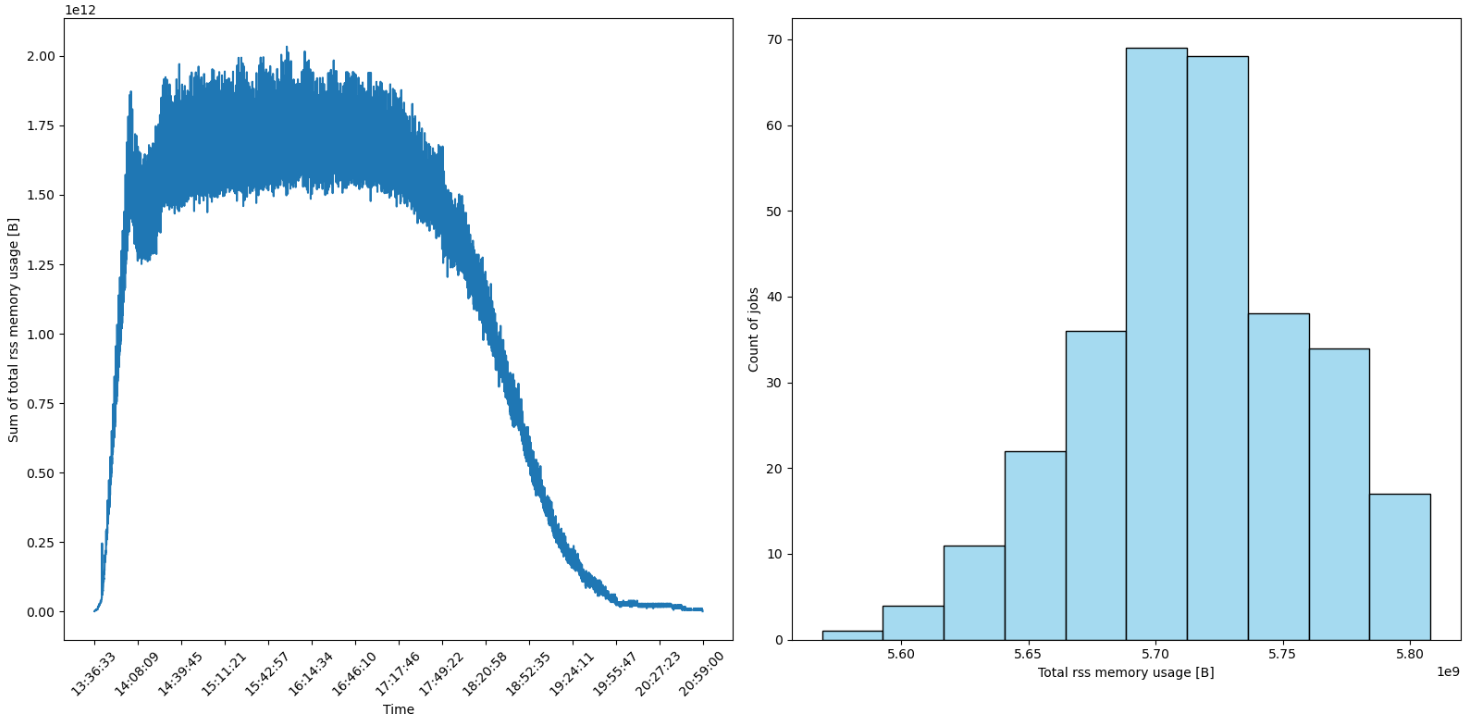}
    \centering
    \caption{ AngioSupport rss memory consumption: total usage during execution (left) and distribution of maximum rss memory usage by jobs (right). }

    \label{fig:grid}
\end{figure}

The distribution reveals that the maximum rss memory consumption by jobs is quite uniform, varying between 5.5 and 5.9 GB, which enables planning of multiple runs per node - depending of its total available memory.

\section{Conclusion}
\label{sec:conc}
We have presented a~solution for end-to-end observability of scientific applications in HPC systems. We have identified the main challenges for such a~solution including instrumentation and propagation of the observability context and collection of application-level metrics. To
address insufficient flexibility of existing visualization tools, we have proposed a~data analysis environment based on JupyterHub and DataFrames. The solution was evaluated using two medical scientific pipelines running on a HPC cluster.


\section*{Acknowledgment}


This publication is partly supported by the EU H2020 grants Sano (857533), ISW (101016503) and by the Minister of Science and Higher Education "Support for the activity of Centers of Excellence established in Poland under Horizon 2020" number MEiN/2023/DIR/3796. 

We gratefully acknowledge Polish HPC infrastructure PLGrid (Cyfronet) within computational grant no. PLG/2023/016227.



%





\end{document}